\documentstyle[12pt,aasms4]{article}

\begin{document}

\title{Polarized CO emission from molecular clouds}
\author{J.S. Greaves, W.S. Holland and P. Friberg}
\affil{Joint Astronomy Centre, 660 N. A`oh\={o}k\={u} Place, 
University Park, Hilo, HI~96720 
\\
{\em jsg@jach.hawaii.edu, wsh@jach.hawaii.edu, friberg@jach.hawaii.edu}}
\and
\author{W.R.F. Dent}
\affil{Royal Observatory, Blackford Hill, Edinburgh EH9 3HJ, UK 
\\
{\em dent@roe.ac.uk}}

\begin{abstract} 

Linearly polarized, non-masing, rotational lines have been detected for
the first time in the interstellar medium. This effect occurs in molecular
clouds with a magnetic field, and traces the field direction, offering an
alternative technique to dust emission polarimetry. The line polarization
mechanism is similar to that in masers, but with lower degrees of
polarization ($\sim$~1~\%).  We have detected the effect towards the
Galactic Centre and its surrounding `2 pc ring', and in the molecular
clouds S140 and DR21 (tentatively), in the lines CO J=2--1 and J=3--2, and
$^{13}$CO J=2--1. The deduced magnetic field directions agree well with
previous dust polarimetry results, confirming that the line polarization
is a real effect. This new technique will be useful in sources that are
too faint for dust polarimetry, and can also be used to investigate
3-dimensional morphology of magnetic fields, where the velocity structure
of the clouds is known.

\end{abstract}


\lefthead{Greaves et al.}
\righthead{Polarized CO emission}

\section{Introduction}

Magnetic fields are thought to play an important role in the evolution of
the interstellar medium. However, in dense molecular clouds where stars
form, these fields are very difficult to observe. The normal technique is
to measure the linearly polarized emission from dust grains, since the
grains are partially magnetically aligned, and have their maximum emission
along the long axes, perpendicular to the field. This technique has been
used successfully at millimetre to far-infrared wavelengths (see
Hildebrand 1996 for a review). 

Alternatively, it has been predicted that the emission from rotating
molecules can be polarized (Goldreich \& Kylafis 1981), and this effect
could also be used to trace magnetic fields. Rotating molecules have a
magnetic moment which interacts with the field, splitting the rotational
quantum levels ({\bf J}) into magnetic sub-levels ({\bf M}). Only
transitions where {\bf M} changes can contribute to emission polarized
perpendicular to the field. Then if the cloud is anisotropic, population
imbalances of the {\bf M} levels arise, and radiation escape probabilities
vary with direction (Goldreich \& Kylafis 1981). Photons with some
particular polarization can escape more readily than others, resulting in
different line intensities for the directions perpendicular and parallel
to the field, and this produces linearly polarized spectra. Typical
polarization levels are predicted to be $\sim$~1~\% (Deguchi \& Watson
1984).

In principle, this effect is an ideal magnetic field tracer, since molecular
lines such as those of CO are bright in many clouds, and a 1~\% modulation
should be detectable. However, previous experiments have been unsuccessful
(Wannier, Scoville \& Barvainis 1983; Barvainis \& Wootten 1987; Lis et al.
1988; Glenn, Walker \& Jewell 1997). The early experiments suffered from
instrumental problems, and some nominal detections were thus thought to be
untrustworthy (Wannier, Scoville \& Barvainis 1983). More recently, improved
receiver performance has lead to much lower polarization limits, such as
p(HCO$^+$~J=1--0) $<$ 0.3~\% in DR21 (Glenn, Walker \& Jewell 1997), below
predicted levels (Lis et al.  1988). This suggests either that the theory is
in error, or that special conditions are required for substantial line
polarization to occur in clouds. The latter hypothesis is supported by the
theoretical work of Deguchi \& Watson (1984), who found that p is maximised
for line optical depth $\tau \sim$~1, and collisional and radiative
transitional rates roughly equal. We have therefore searched for
polarization in CO lines, in moderately dense molecular environments, where
these conditions may be met. In this Letter, we present the first line
polarization detections, in a range of molecular clouds and three
independent spectral lines. These results are complementary to very recent
detections of line polarization in the circumstellar envelope of the AGB
star IRC+10216 (Glenn et al. 1997), and in compact cores surrounding OMC1/
IRc2 (BIMA 1.3 mm interferometry; D. Crutcher and R. Rao, priv. comm.).

\section{Observations}

The observations were made in the period July 1995 to July 1997, at the
James Clerk Maxwell Telescope located on Mauna Kea, Hawaii. A rotating
quartz half-wave plate (Murray et al. 1992) was mounted in front of the
single-polarization facility heterodyne receiver RxA2 (Davies et al.
1992). The data were taken by stepping the waveplate at 22.5$^{\circ}$
intervals around 360$^{\circ}$, thus rotating the source plane of
polarization with respect to that accepted by the receiver, and recording
a spectrum at each position. Sky emission was subtracted by switching to
an off-source position up to a degree away, and the integrated intensities
of the lines were then fitted for the polarization modulation using a
least-squares procedure (Nartallo 1995). Instrumental polarization (IP)
was subtracted and the residual signal was corrected for source
parallactic angle, and a rotation corresponding to the receiver position. 
(The absolute angle of the waveplate's fast-axis had been established
previously with a polarizing grid). The final percentage polarization and
error in the position angle were corrected for bias effects (due to the
vector nature of p), as described by Wardle \& Kronberg (1974). Polarized
spectra were also constructed, using a direct subtraction method (e.g.
Wannier, Scoville \& Barvainis 1983), and IP and rotation corrections as
above.

Integration times used were generally 16 seconds per waveplate position,
sufficiently short that sky variations had only a small effect on the
signal. Some automatic despiking was used in the analysis, removing
typically 1--2 of the 16 fitted points in order to improve the results in
a chi-squared test. For one source (SW in CO J=2--1, see below), an
empirical criterion was also used to reject some cycles completely, where 
about half of the noise estimates per cycle showed anomalously high
values, indicating serious sky instability. Between 5 and 17 complete
waveplate cycles per source were made to detect the polarization, with
signal-to-noise ratios of typically 3.5 ($>$ 99 \% confidence levels).
This is sufficient to determine the position angle error $\delta\theta$ to
$\pm$ 8$^{\circ}$ (e.g. Wardle \& Kronberg 1974).

An important aspect of the observations is the removal of instrumental
polarization. An incoming unpolarized signal will become polarized due to
effects of the telescope optics, and transmission through the woven JCMT
windblind, which has a spacing comparable to millimetre wavelengths. We
measured this IP using Mars, Saturn and Jupiter, since the planets have
very low intrinsic degrees of polarization (Clemens et al. 1990). The IP
levels were 0.8--1.1~\%, with measured errors of 0.15--0.3~\%. For the
reduction in this paper we have mainly used Saturn, which had an apparent
diameter of 17--19$''$ during our observations, and was thus well matched
to the JCMT beam (21$''$ FWHM at the CO J=2--1 frequency of 230 GHz).
During the CO J=3--2 observations at 345 GHz, Saturn was not available,
and Jupiter was used instead (45$''$ diameter). The IP's were found to
depend only slightly on frequency within the passband ($\delta$IP =
$\pm$~0.0--0.3~\%), and varied little with source elevation ($\delta$IP
$\sim \pm$~0.15~\%). The IP's subtracted were for the main beam only, but
sidelobe instrumental polarization was found to be negligible.
Measurements using Jupiter at 230 GHz showed that the effective off-axis
IP contributed a residual of around 0.1~\%, with the planet offset just
outside the main beam, and covering a region 45$''$ in diameter in
sidelobes of $\sim$ 1~\% amplitude relative to the main beam.

We also investigated various instrumental effects that can mimic a
polarization signal, including calibration changes, reflection modulations
and beam-shape distortions. Calibration was done using hot, cold and sky
loads at the start of each waveplate cycle, and an additional calibration,
using the off-position as the sky signal, was done for each individual
spectrum. The intensity calibration is therefore very accurate. Waveplate
reflections, which can modulate the signal as the plate rotates, were
minimised by mounting the polarimeter module at a slight angle to the
incoming beam. Finally, an elliptical beam could cause signal modulations, as
the waveplate movement effectively rotates the beam over the source. Bright
regions at the edges of the beam could then be included in the signal for
some waveplate angles. However, we have estimated this effect, using as an
example a map around our SW Galactic Centre point (see below), and for a 5~\%
beam ellipticity, the effective `polarization' introduced is $<$~0.05~\%.

Instrumental polarization can thus be subtracted quite reliably, to a limit
of about $\delta$IP $\sim$~0.3~\%. This uncertainty arises mainly from the
choice of planet to use for the IP, since it is difficult to match the
planet and source sizes exactly. Our line polarimetry technique will
therefore only give reliable position angles, and hence field directions,
when the source polarization is significantly larger than $\sim$~0.3~\%.

\section{Results}

The predicted polarization level is maximised if the source has well-ordered
magnetic and velocity fields (Goldreich \& Kylafis 1981; Lis et al. 1988),
so we searched for sources which met these criteria. An ideal target is the
`2 pc ring' around the Galactic Centre, which has a roughly linear magnetic
field and constant rotational velocity (Hildebrand et al.  1993; Sutton et
al. 1990). Two positions were observed in this ring, and the results for CO
J=2--1 polarization are presented in Table~1. Consistent results were
obtained on two separate runs, with $\delta$p differing at only about the
1.5~$\sigma$ level. The observations were made at different parallactic
angles for the two runs, so the agreement after de-rotation indicates that
we are not just measuring an instrumental effect.

To confirm these detections, we re-observed one of these positions in CO
J=3--2, obtaining a similar position angle. We also looked for
polarization in $^{13}$CO J=2--1 towards Sgr A*, and observed two
molecular clouds to determine if line polarization is also present in
star-formation regions. S140 was detected in $^{13}$CO J=2--1
polarization, and DR21 in CO J=2--1 (tentatively).  These results are
listed in Table~1.

A good test of the reality of the line polarization is the position angle
measured. This can be either parallel or perpendicular to the magnetic field
(Kylafis 1983), and should thus be at 0$^{\circ}$ or 90$^{\circ}$ to the
direction of dust polarization (which is always perpendicular to the field).
For the Sgr A* and 2 pc ring positions, 100~$\mu{m}$ dust polarization has
been detected by Hildebrand et al. (1993), and there is good agreement with
the CO and $^{13}$CO results. The difference in position angle is $2 \pm
8^{\circ}$ for the NE ring position, in the CO J=2--1 line, and $81 \pm
9^{\circ}$ for the SW position, in the CO J=2--1 and J=3--2 lines combined.
(The CO J=3--2 result alone has a larger angle deviation, but this could be
explained by small-scale field structure, since the beam size is
$\approx$~2.5 times smaller than for the 100~$\mu{m}$ data.) The line
polarization is found to be perpendicular to the field at the NE point, and
parallel towards SW, a difference which is probably related to viewing angle
effects (Kylafis 1983), as the points lie at different ring radii.

For Sgr~A*, the projected ring velocity is $\approx$~0~km s$^{-1}$, similar
to foreground gas, so the $^{13}$CO polarization was found for the complete
spectrum. The difference in angle with the dust data is $4 \pm 9^{\circ}$,
implying that both of these optically-thin tracers are detecting the same
net magnetic field, for the clouds along the line of sight.

The two molecular cloud sources observed were DR21 and S140-IRS1, which
have previously been observed in 800~$\mu{m}$ dust polarimetry at the JCMT
(Minchin \& Murray 1994; Minchin, Bonif\'{a}cio \& Murray 1996). For DR21,
the CO optical depth is very high ($\tau \sim 7$ from comparison with a
$^{13}$CO spectrum, and assuming $^{12}$C/$^{13}$C = 53; Langer \&
Penzias 1990). It is therefore not surprising that the measured p is
small and only marginally detected, since both polarizations tend to
saturate (Goldreich \& Kylafis 1981). Also, since only the front parts of
the cloud will be detected in CO, the estimated position angle is not the
same as in the dust data (Table~1). In contrast, for the more optically
thin $^{13}$CO emission in S140-IRS1, there is very good agreement in
position angles ($\delta\theta = 6 \pm 10^{\circ}$).

\section{Discussion}

Polarization has been detected at five points in star-formation regions
and around the Galactic Centre, in a total of three spectral lines. The
degree of polarization ranges from 0.5--2.4~\%, and is at about the level
predicted theoretically. For example, Deguchi \& Watson (1984) modelled
the polarization of the J=2--1 transition of CO, assuming simple linear
magnetic fields and velocity gradients, and found p up to 0.5--4.5~\% for
gas densities of $\sim 10^4$ cm$^{-3}$. Also, the position angles found
from the line observations are in excellent agreement with those derived
previously from dust polarimetry. The combination of these factors
indicates that real detections of line polarization have been made. 

The results have shown that polarized CO and $^{13}$CO lines are good tracers of
magnetic fields, under suitable conditions. These are chiefly that the optical
depth should be $\sim$~1, and that the collisional and radiative rates for the
transition should be comparable (Goldreich \& Kylafis 1981). For the CO J=2--1
line, the latter condition corresponds to gas densities of $\sim 10^4$ cm$^{-3}$
at temperatures of 30 K, which are realistic parameters for core regions of
giant molecular clouds. In the particular case of the Galactic centre ring, the
estimated n$_{H_2}$ is $2 \times 10^4$ cm$^{-3}$ (Sutton et al. 1990;  Zylka,
Mezger \& Wink 1990), and the ring optical depth is estimated at 1.3--1.6 (using
the $^{13}$CO/CO line ratio and an isotopic ratio of 25, Wilson \& Matteucci
1992; Langer \& Penzias 1990).  Under these conditions, p(CO J=2--1) should be
$\leq$~1.5~\% (Deguchi \& Watson 1984), and our measured p(CO) are $\approx$
0.7--1.7~\%, in good agreement. In some circumstances, other effects could
contribute to the polarization, such as selective excitation of the sub-levels
by IR-pumping (Morris, Lucas \& Omont 1985), but this will not be significant
for general cloud material, and even for the strongly centrally-illuminated
Galactic Centre region, we do not see the predicted radial or tangential
polarization directions in the CO data (Table 1).

The polarization level falls below detectable limits (p comparable to
0.3~\%)  for high $\tau$, as for example in DR21 (Table 1). We have also
searched for CO J=3--2 polarization in more optically thick regions of the
Galactic Centre ring, and found p typically only $\sim$~0.4~\%. Thus,
previous non-detections of line polarization may have been partly due to
high optical depths in the molecular cores observed. Density is also
critical, as at high n$_{H_2}$, collisions scramble the sublevel level
populations, and this can explain, for example, low polarization in the
very dense NH$_3$ core in OMC1 (p~$<$~1.1~\%, Barvainis \& Wootten 1987).
More recently, Glenn, Walker \& Jewell (1997) have measured a low
polarization in the DR21 core in the HCO$^+$ J=1--0 line. They found p
$\sim 0.24 \pm$ 0.04 \% (below the level of possible instrumental effects)
and this low level may be due to the high core optical depths and/or
densities. The CO lines of moderate optical depth therefore appear to be
the most suitable for spectropolarimetry, as they trace moderate density
material in more extended regions of molecular clouds.

A particular advantage of line polarimetry is that the magnetic field
components of different clouds along the line of sight can be separated.
This is illustrated in Figure~1, showing the polarized CO J=2--1 spectra
towards the NE- and SW-ring positions. The bottom two spectra are the Stokes
parameters Q(v) and U(v) (see e.g. Wannier, Scoville \& Barvainis 1983),
representing orthogonal components of the polarization, and the top plot
shows the total intensity spectrum I(v) (on a 1~\% scale). The percentage
polarization is found from p = I$^{-1} \times \sqrt{}$(Q$^2$ + U$^2$), and the
position angle is given by $\theta$ = 1/2 tan$^{-1}$(U/Q). It can be seen
that other sources, such as the well-known `+20 km s$^{-1}$' and `+50 km
s$^{-1}$' clouds (e.g. Zylka, Mezger \& Wink 1990) are also polarized, but
have Q and U components of different magnitude and sign to the ring,
implying different field directions. (Note that the exact polarization
parameters for these clouds are uncertain, as they are spatially extended 
so the subtraction of the Saturn IP may not be appropriate.)

Figure~1 illustrates that net polarization for the whole spectrum will not
represent any of the individual magnetic fields. This is a problem for
dust polarimetry, where the clouds cannot be separated (although in the
case of the `2 pc' ring, there is little dust emission from foreground
clouds, Hildebrand et al. 1993). The particular advantage of line
polarimetry is that field directions for clouds at different velocities
can be determined separately. Thus, this new polarimetry technique
provides substantial information on the 3-D field structure in sources
with velocity discrimination (such as rotation, infall or outflow), even
though the line-of-sight field components cannot be detected.

\section{Conclusions}

Polarized rotational lines have been detected for the first time in molecular
cloud sources. The polarization levels are $\approx$~0.5--2~\%, and the deduced
magnetic field directions agree very well with previous dust polarimetry data.
The advantages of line polarimetry are that field orientation can be found as a
function of velocity, and also that faint sources can be observed, that may not
have significant dust emission. A 90$^{\circ}$ directional ambiguity can be a
problem for some sources, although this may be resolved with sufficient source
information (Kylafis 1983). CO emission is often bright, even where the dust
(sub)millimetre continuum is not detectable, such as in low column density
sources. This new technique thus has potential for mapping the 3-D magnetic
structure of sources such as star-formation regions, protostellar core/outflow
systems and circumstellar envelopes, as well as understanding the Galactic
centre field morphology.

\acknowledgements

We are very grateful to Sye Murray and Ramon Nartallo for useful discussions
regarding the polarimetry hardware and software, and John White for
technical help. The JCMT is operated by the Joint Astronomy Centre, on
behalf of the UK Particle Physics and Astronomy Research Council, the
Netherlands Organisation for Pure Research, and the National Research
Council of Canada.

\newpage

\begin{center}
{\bf Figure Captions}
\end{center}

\figcaption[f1.ps] {Polarization spectra towards the NE (left) and SW (right)
ring positions.  The top spectrum is total intensity I, on a T$_A^*$ antenna
temperature scale, divided by 100. The middle and bottom spectra are the Stokes
parameters Q and U (see text) and the vertical bars represent typical standard
errors in each detected channel. The ring velocities (Serabyn et al. 1986) are
marked by the dashed vertical lines. }

\newpage


\begin{deluxetable}{lccccccc}
\tablenum{1}
\tablewidth{0pt}
\tablecaption{Line polarization results}
\tablehead{
\colhead{Source} & \colhead{date} & \colhead{line} &
\colhead{n$_{\rm cycles}$} & \colhead{p (\%)} & \colhead{$\theta$
($^{\circ}$)} & \colhead{$\theta_{\rm dust}$ ($^{\circ}$)} &
\colhead{$\delta\theta$ ($^{\circ}$)} } 
\startdata
Sgr A*  & 7/97 &$^{13}$CO J=2--1& 5 & 2.39 $\pm$ 0.81 & 91 $\pm$ 9 & 
  95 $\pm$ 3 & 4 $\pm$ 9 \nl
NE-ring & 7/95 & CO J=2--1 & 10 & 0.89 $\pm$ 0.23 & 74 $\pm$ 7 & & \nl
        & 7/97 & CO J=2--1 &  7 & 0.54 $\pm$ 0.12 & 58 $\pm$ 6 & & \nl 
        &95\&97& CO J=2--1 & 17 & 0.73 $\pm$ 0.15 & 69 $\pm$ 6 & 
  71 $\pm$ 6 & 2 $\pm$ 8 \nl
SW-ring & 7/95 & CO J=2--1 &  5 & 2.15 $\pm$ 0.51 & 20 $\pm$ 7 & & \nl
        & 7/97 & CO J=2--1 &  3 & 0.93 $\pm$ 0.61 & 14 $\pm$ 20& & \nl
        &95\&97& CO J=2--1 &  8 & 1.74 $\pm$ 0.43 & 18 $\pm$ 7 & 
  112 $\pm$ 4 & 86 $\pm$ 8 \nl
SW-ring & 7/96 & CO J=3--2 & 11 & 1.33 $\pm$ 0.40 & 44 $\pm$ 8 & 
  112 $\pm$ 4 & 68 $\pm$ 9 \nl
SW-ring &95--97& CO J=2--1,3--2 & 19 & --- & 31 $\pm$ 8 & 112 $\pm$ 4 & 
  81 $\pm$ 9 \nl
&&&&&&&\nl
S140-IRS1& 6/97& $^{13}$CO J=2--1 & 5 & 0.77 $\pm$ 0.23 & 15 $\pm$ 8 & 
  21 $\pm$ 6 & 6 $\pm$ 10 \nl
DR21    & 7/95 & CO J=2--1 & 7.5 & (0.23 $\pm$ 0.32) & (13 $\pm$ 38) &&\nl
        &12/96 & CO J=2--1 & 6   & 0.74 $\pm$ 0.27 & 146 $\pm$ 10 & & \nl
        &95\&96& CO J=2--1 & 13.5& 0.51 $\pm$ 0.22 & 142 $\pm$ 11 & 
  17 $\pm$ 4 & 55 $\pm$ 12 \nl 
\enddata
\tablecomments{Results given are the average polarization p and position
angle $\theta$ after the number of waveplate cycles indicated. (For the
combined data in two lines at the SW position, normalised Stokes
parameters were used.) The dust polarimetry position angles $\theta_{\rm
dust}$ are from 100~$\mu{m}$ observations (Galactic Centre positions,
Hildebrand et al.  1993), and 800~$\mu{m}$ observations at the JCMT
(star-formation regions, Minchin \& Murray 1994; Minchin, Bonif\'{a}cio \&
Murray 1996). The difference in the dust and spectral line position angles
is given by $\delta\theta$. The positions described as NE and SW are
offset by $\delta$R.A.,$\delta$Dec. = +56,+93 and --56,--56, respectively,
from Sgr~A*; other position co-ordinates are given in the dust polarimetry
references.} 
\end{deluxetable}


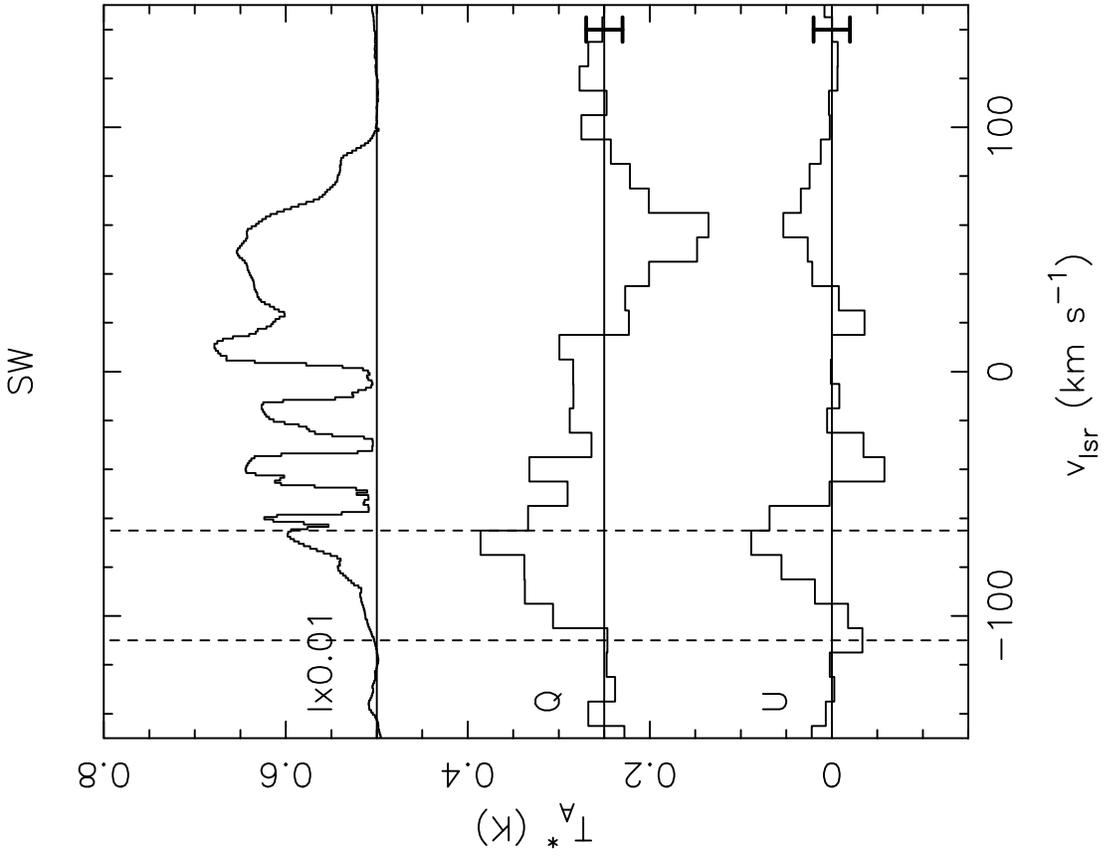
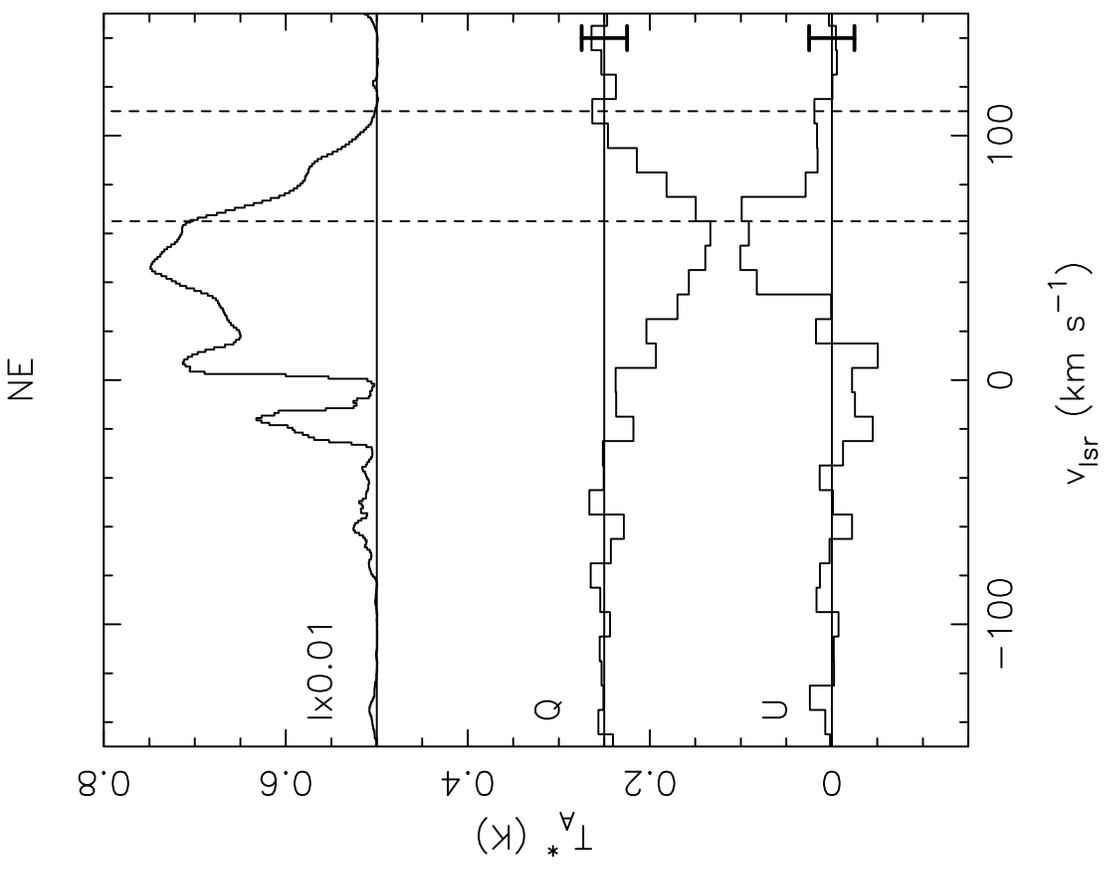
\end{document}